\documentclass{myaa}

\usepackage[varg]{txfonts}

\usepackage{natbib}
\usepackage{graphicx}
\usepackage{epstopdf}
\usepackage{amssymb}
\usepackage{float}
\begin{document}

\title{Preliminary model of the outer disk of RU Lup presently showing only four dark gaps}
%%%  $\boldsymbol{G}$

\author{Dimitris M. Christodoulou\inst{1,2}  
\and 
Demosthenes Kazanas\inst{3}
}

%%%$^{2}$\footnotemark[1]
\institute{
Lowell Center for Space Science and Technology, University of Massachusetts Lowell, Lowell, MA, 01854, USA.\\
\and
Dept. of Mathematical Sciences, Univ. of Massachusetts Lowell, 
Lowell, MA, 01854, USA. \\ E-mail: dimitris\_christodoulou@uml.edu\\
\and
NASA/GSFC, Laboratory for High-Energy Astrophysics, Code 663, Greenbelt, MD 20771, USA. \\ E-mail: demos.kazanas@nasa.gov \\
}

%\date{Received~~2018 month day; accepted~~2019~~month day}

\def\gsim{\mathrel{\raise.5ex\hbox{$>$}\mkern-14mu
                \lower0.6ex\hbox{$\sim$}}}

\def\lsim{\mathrel{\raise.3ex\hbox{$<$}\mkern-14mu
               \lower0.6ex\hbox{$\sim$}}}

\abstract{
Recent ALMA/DSHARP observations reveal that the disk of RU Lup shows presently only four dark gaps at large radii in which planets are presumably already orbiting. This gap arrangement is quite similar to the current structure of the outer planets in our solar system. Although there may be more undetected planets forming in the inner disk of RU Lup, the information obtained from the current observations prompted us to fit preliminary models with an isothermal oscillatory density profile to the outer disk of RU Lup. The best-fit model confirms that several physical properties of the disk of RU Lup are comparable to those of our solar nebula to within factors of 1.0-2.3; but there are some differences as well: central density ($\rho_0$), centrifugal support ($\beta_0$), and core angular velocity ($\Omega_0$) in RU Lup are smaller by factors of 5.5, 8.0, and 18, respectively, whereas the inner core radius ($R_1$) of RU Lup is 25 times larger. The smaller values found for RU Lup are linked to the (apparently gap-empty) inner 10 AU of its disk,  a region that is presently inadequately resolved.}


\authorrunning{ }
\titlerunning{Physical properties of the disk of RU Lup}

\maketitle

\section{Introduction}\label{intro}

Recent ALMA/DSHARP observations of young ($\sim$0.1-10 Myr) protostellar gaseous disks have revolutionized our understanding of the structures and dynamics of these systems \citep{vdm19}. The most prominent features seen for the first time are multiple dark gaps in the disks that were presumably carved out by orbiting protoplanets \citep{alm15,and16,rua17,lee17,lee18,mac18,ave18,cla18,kep18,guz18,ise18,zha18,dul18,fav18,har18,hua18,per18,kud18,lon18,pin18}.      Motivated by these high-resolution ($\sim$1-5~AU) observations, we constructed equilibrium models of such planet-forming disks in the early ($<0.1$ Myr) isothermal phase of their evolution. Previously, we modeled our own solar nebula and the ALMA-observed disks of AS 209 and HL Tau \citep[][respectively]{chr19a,chr19b}. The main conclusion from these investigations was that these systems follow a ``bottom-up'' hierarchical formation scenario according to which the protostars form long after their planets and the regular moons have already been formed.

In this paper, we apply the same theoretical model of early planet formation to the outer disk of RU Lup \citep{hua18}, a young protostellar system that exhibits striking similarities to our much older solar system (Table~\ref{table1}). Our goal is to quantify similarities and physical differences between these two systems. The analytic (intrinsic) and numerical (oscillatory) solutions of the isothermal Lane-Emden equation \citep{lan69,emd07} with differential rotation, and the resulting model of the midplane of the gaseous disk have been described in detail in \cite{chr19a}. In \S~\ref{models2}, we describe our best-fit model of the disk of RU Lup and we compare its physical properties to those of the solar nebula, as these were derived from the current arrangement of the planets in our solar system. In \S~\ref{disc}, we summarize these preliminary results.

\section{A Physical Model of the RU Lup Protostellar Disk}\label{models2}

As in previous related work, the numerical integrations that produce oscillatory density profiles were performed with the \textsc{Matlab} {\tt ode15s} integrator \citep{sha97,sha99} and the optimization used the Nelder-Mead simplex algorithm as implemented by \cite{lag98}. This method (\textsc{Matlab} routine {\tt fminsearch}) does not use any numerical or analytical gradients in its search procedure which makes it extremely stable numerically, albeit somewhat slow and computationally expensive. The boundary conditions for the oscillatory density profiles are, as usual, $\tau(0)=1$ and $[d\tau/dx](0)=0$, where $\tau$ and $x$ are the dimensionless values of the density and the radius, respectively.

\subsection{Best-Fit model of RU Lup}\label{model1}

The radii of the four dark gaps of RU Lup are listed in Table~\ref{table1} and they are compared to the semimajor axes of the outer planets in our solar system. It is obvious that these two arrangements are quite similar. We note that Jupiter is missing from the list, and this is telling. We have not found any acceptable model of RU Lup in which the first density peak can reach out close to the location of the innermost gap D14 (at 14 AU). We were forced to leave the first density peak (at $R=7.29$ AU) empty, and then we fitted peaks \#2-5 to the four observed gaps. We have encountered a similar problem in the modeling of the moons of Neptune \citep{chr19c}, where we had to leave the two innermost density peaks empty. Our choice is understandable in both cases: the inner core of RU Lup is not adequately resolved and there could be more protoplanets growing in it; and the satellites of Neptune have been severely disturbed by the retrograde capture of Triton deep inside the planet's protoplanetary disk.

\begin{figure}
\begin{center}
    \leavevmode
      \includegraphics[trim=0.2 0.2cm 0.2 0.2cm, clip, angle=0, width=10 cm]{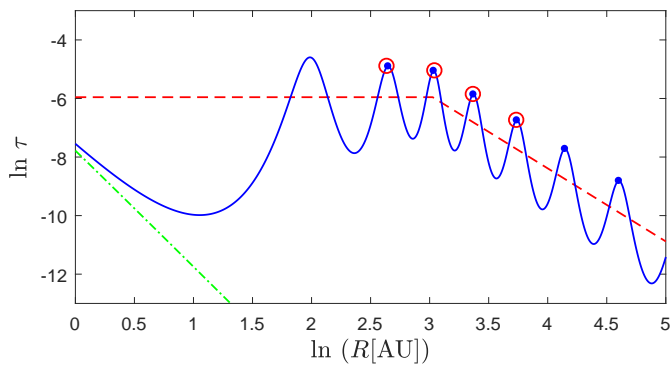}
      \caption{Equilibrium density profile for the midplane of RU Lup disk that has already formed at least four annular dark gaps (presumably protoplanets) \citep{hua18}. The best-fit parameters are $k=-1.5$, $\beta_0=0.0509$, and $R_1=20.6$~AU. The radial scale length of the disk is $R_0=0.051$~AU. The Cauchy solution (solid line) has been fitted to the dark gaps of RU Lup (Table~\ref{table1}) so that its density maxima (dots) correspond to the observed orbits of the protoplanets (open circles). The first density peak at $R=7.29$ AU had to be left empty in order to achieve a good-quality fit. The density peak corresponding to the location of the fourth maximum was scaled to the observed radius of $R_4=29.1$~AU. The mean relative error of the fit is 1.0\%. The intrinsic analytical solution (dashed line) and the nonrotating analytical solution (dash-dotted line) are also shown for reference. 
\label{fig1}}
  \end{center}
  %\vspace{-1.35cm}
\end{figure}

In Figure~\ref{fig1}, we show the best optimized fit to the dark gaps of RU Lup. In the models, we have used only three free parameters ($k$, $\beta_0$, and $R_1$); there is no need to utilize an outer flat-density region because the observed gaps do not extend very far out. In fact, the next outer density peak along the power-law gradient lies at 63.2 AU, which falls well outside the maximum observed disk radius of $R_{\rm max}\simeq 50$ AU. With the assumption of leaving the innermost density peak empty (that would correspond roughly to Jupiter's orbit in our solar nebula), the mean relative error of the best fit is 1.0\%, indicating that this is a high-quality fit. 

Objections may still be raised about this model of the outer disk of RU Lup: we have effectively used 3 free parameters to fit the locations of only 4 dark gaps. We believe that these objections have no merit because in the original modeling effort with 4 free parameters, 4 gaps, and without any empty inner peaks, we utterly failed to find an acceptable model. Thus, a number of free parameters comparable or equal to the number of suspected planets does not guarantee that a good-quality model will be found.

\begin{table}
\caption{Dark gaps in RU Lup \citep[Table 1 in][]{hua18} versus solar-system planets}
\label{table1}
\begin{tabular}{ll|ll}
\hline
Gap    & RU Lup     & Planet    &  Solar System \\
Name & $R~(AU)$ & Name  &  $R~(AU)$ \\
\hline
D14  &   14      &   Saturn   &   9.54 \\
D21  &   21      &   Uranus  &   19.2 \\
D29  &  29.1    &   Neptune  &  30.1 \\
D42  &  42       &   Pluto   &   39.5 \\

\hline
\end{tabular}
\end{table}

\begin{table*}
\caption{Comparison of the protostellar disk models of RU Lup and our solar nebula}
\label{table2}
\begin{tabular}{llll}
\hline
Property & Property & RU Lup & Solar Nebula \\
Name     & Symbol (Unit) & Best-Fit Model & Best-Fit Model \\
\hline
Density power-law index & $k$                                          &   $-1.5$  	     & $-1.5$    \\
Rotational parameter & $\beta_0$                                &    0.0509     	   &  0.41   \\
Inner core radius & $R_1$ (AU)                              &   20.6  	       			&  0.82    \\
Scale length & $R_0$ (AU)                               &       0.051   	   &  0.022   \\
Equation of state & $c_0^2/\rho_0$ (${\rm cm}^5 {\rm ~g}^{-1} {\rm ~s}^{-2}$) & $4.9\times 10^{17}$ & $9.0\times 10^{16}$    \\
Minimum core density for $T=10$~K, $\overline{\mu} = 2.34$ & $\rho_0$ (g~cm$^{-3}$)         &    $7.3\times 10^{-10}$   			&   $4.0\times 10^{-9}$   \\
Isothermal sound speed for $T=10$~K, $\overline{\mu} = 2.34$ & $c_0$ (m~s$^{-1}$) & 188 & 188 \\
Jeans gravitational frequency & $\Omega_J$ (rad~s$^{-1}$)    &    $1.8\times 10^{-8}$ & $4.0\times 10^{-8}$    \\
Core angular velocity & $\Omega_0$ (rad~s$^{-1}$)    &    $8.9\times 10^{-10}$ 	& $1.6\times 10^{-8}$    \\
Core rotation period & $P_0$ (yr)                                 &    224 	   			&  12   \\
Maximum disk size & $R_{\rm max}$ (AU)                &    50 	   			&   50  \\
\hline
\end{tabular}
\end{table*}

\subsection{Comparison between RU Lup and the solar nebula}\label{comp}

The physical properties of the best-fit models of RU Lup and the solar nebula \citep{chr19a} are listed in Table~\ref{table2}. We see several similarities in the values of $k$, $R_0$, $\Omega_J$, and $R_{\rm max}$; but also four noticeable differences: 
\begin{enumerate}
\item The centrifugal support of RU Lup against self-gravity ($\beta_0$) is 8 times smaller, making this disk exceptionally stable to dynamical nonaxisymmetric instabilities for millions of years to come \citep{chr95}.
\item The inner core radius ($R_1$) of RU Lup very large; it is 25 times larger that the core of the solar nebula. This could very well be because the inner 10 AU of RU Lup are not currently adequately resolved, and they could be hiding more orbiting planets, especially smaller planets that are not as efficient in carving too wide dark gaps.
\item The central density ($\rho_0$) of RU Lup is 5.5 times smaller. This is because $\rho_0\propto 1/R_0^2$. The ratio of scale lengths is only $\approx$2.3, but its square takes a larger value.
\item The angular velocity of the core ($\Omega_0$) of RU Lup is 18 times smaller, so this large core rotates very slowly. This is a direct consequence of the large size of the core of RU Lup because the two models have similar values of the Jeans gravitational frequency $\Omega_J$ (to within a factor of 2.2) and disk size ($R_{\rm max}\simeq 50$ AU), which imply that the disks have comparable weights.
\end{enumerate}

The power-law index of $k=-1.5$ found in both models strengthens the view that the two disks are similar in structure. This is also approximately the power-law index of the radial surface density profile (i.e., $\Sigma(R)\propto R^{-1.5}$) which is not too different than those observed in other protostellar systems \citep{and07,hun10,lee18} and deduced for our solar nebula in the seminal work of \cite{wei77}.

\section{Summary}\label{disc}

In \S~\ref{models2}, we presented our best-fit isothermal differentially-rotating protostellar model of the young system RU Lup, recently observed by ALMA/DSHARP \citep{hua18}. RU Lup shows only four dark gaps in its outer disk, an arrangement of presumed protoplanets that is quite similar to the outer planets in our solar system (see Table~\ref{table1}). The best-fit model is shown in Figure~\ref{fig1} and a comparison of its physical properties versus those of the solar nebula is shown in Table~\ref{table2}.

The two disks appear to be similar in structure and they both obey the Titius-Bode rule (Table~\ref{table3}). We found four differences in their physical properties, but all of which derive from the unresolved inner 10 AU inside the enormous core of RU Lup: In addition to the pronounced difference in core sizes (by a factor of 25), parameters $\rho_0$, $\beta_0$, and $\Omega_0$ are smaller in RU Lup by factors of 5.5, 8.0, and 18, respectively.

\begin{table*}
\caption{Titius-Bode Geometric Progression in RU Lup}
\label{table3}
\begin{tabular}{ll|ll|ll}
\hline
          &                &  Density     &                 & Geometric   &   \\
Gap    & RU Lup     & Maximum  &                 & Progression &  \\
Name & $R$         &  $d$           &  $d$-Error &  $(d_{i-1}\cdot d_{i+1})^{1/2}$  &  p-Error \\
          &   (AU)      &   (AU)         & (\%)        &    (AU)                                          & (\%) \\
\hline
D14  &   14            &  14.1 & +0.71  &            &      \\
D21  &   21            &  20.8 &  $-0.95$ &  20.3   &  $-2.4$    \\
D29  &  29.1          &  29.1 &   0 &  29.6   &   +1.7   \\
D42  &  42            &   42.0 &   0 &  42.9   &   +2.1   \\
        &                  &   63.2 &   &  64.8   &   +2.5    \\
        &                  &   99.8 &   &    &      \\

\hline
\end{tabular}

Notes.---The $d$-Error is the deviation of each density maximum from the corresponding observed D-gap. The p-Error is the deviation of each geometric progression distance, ($d_{i-1}\cdot d_{i+1})^{1/2}$, from the corresponding density maximum $d$.
\end{table*}

Furthermore, the disk of RU Lup appears to be extremely stable and long-lived ($\beta_0 = 0.0509$), so we believe that the system will continue to evolve in a nonviolent fashion for millions of years to come (i.e., there will be no planet-planet resonant interactions,\footnote{We checked and the currently observed dark gaps do not show any low-order mean-motion resonances, just as the planets in our solar system.} no planet-gas disk interactions, and no orbital migrations to disturb or destroy this young protoplanetary system). We also believe that the observed well-organized planet formation will conclude within the observed dark gaps long before the gaseous disk disappears and the central star finally emerges. These results continue to support strongly a ``bottom-up'' formation scenario in which planets form first in $< 0.1$ Myr \citep{gre10,har18}, followed by the formation of their central stars after the gaseous disks have been cleared out by the ongoing accretion and mass-loss processes.

\iffalse
\section*{Acknowledgments}
%We thank anonymous referees for suggestions that clarified considerably the presentation of these ideas. 
NASA support over the years is gratefully acknowledged.
\fi

%\label{lastpage}

\end{document}